\documentclass[11pt]{article}
\usepackage{a4wide}
\usepackage{color}

\usepackage[bookmarks=true,bookmarksnumbered=true,setpagesize=false]{hyperref}

\usepackage{enumerate}
\usepackage{amsmath,amssymb,amsthm}
\usepackage{textcomp,mathcomp} 

\usepackage{mathtools}

\usepackage{graphicx}
\usepackage{braket}
\usepackage{url}

\usepackage{mathrsfs}

\newcommand{\h}{\hat}

\newcommand{\tx}{\text}

\usepackage{comment}

\newcommand{\bp}{\begin{pmatrix}}
\newcommand{\ep}{\end{pmatrix}}
\newcommand{\bb}{\begin{bmatrix}}
\newcommand{\eb}{\end{bmatrix}}

\DeclareMathOperator{\E}{E}
\DeclareMathOperator{\V}{V}

\newcommand{\df}{\text{d}}

\newcommand{\bs}{\boldsymbol}

\newcommand{\bh}[1]{\boldsymbol{\hat{#1}}}

\newcommand{\al}[1]{\begin{align}#1\end{align}}

\newcommand{\ol}{\overline}

\newcommand{\autospace}{%
  \mathchoice%
    {\!}
    {\!}
    {}
    {}
}
\newcommand{\fn}[1]{\autospace\paren{#1}} 
\newcommand{\fnl}[1]{\autospace\sqbr{#1}} 

\newcommand{\paren}[1]{\left(#1\right)}
\newcommand{\pn}[1]{\left(#1\right)}
\newcommand{\sqbr}[1]{\left[#1\right]}

\newcommand{\nn}{\nonumber\\}

\newcommand{\sr}{\stackrel}

\newcommand{\red}[1]{{\color[cmyk]{0,0.8,1,0}#1}}
\newcommand{\green}[1]{{\color[cmyk]{0.97,0,0.75,0}#1}}
\newcommand{\blue}[1]{{\color[cmyk]{1,0.5,0,0}#1}}

\newcommand{\trv}{\red} 
\newcommand{\erv}{\blue} 
\newcommand{\tv}{\red} 
\newcommand{\ev}{\blue} 
\newcommand{\fv}{\green} 

\usepackage{fancybox}

\usepackage[bbgreekl]{mathbbol}
\usepackage{amsfonts}

\DeclareSymbolFontAlphabet{\mathbb}{AMSb}
\DeclareSymbolFontAlphabet{\mathbbl}{bbold}

\usepackage{amsthm}
\theoremstyle{definition}

\newcommand{\ov}{\over}

\newcommand{\tf}{\textsf}

\newcommand{\mc}{\mathcal}

\newcommand{\anotherautospace}{%
  \mathchoice%
    {\;}
    {\;}
    {\,}
    {\,}
}
\newcommand{\md}{\anotherautospace\middle|\anotherautospace}

\usepackage{slashed}

\newcommand{\sD}{s}
\newcommand{\rD}{r}
\newcommand{\dD}{d}
\newcommand{\lamD}{\lambda}
\newcommand{\kD}{k}

\begin{document}
\title{Implementing Errors on Errors: Bayesian vs Frequentist}
\author{Satoshi Mishima\thanks{E-mail: \tt mishima@saitama-med.ac.jp}\ \mbox{} and Kin-ya Oda\thanks{E-mail: \tt odakin@lab.twcu.ac.jp}\bigskip\\
\normalsize\it
$^*$Department of Liberal Arts, Saitama Medical University, Saitama 350-0495, Japan\\
\normalsize\it
$^\dagger$Department of Information and Mathematical Sciences,\\
\normalsize\it
Tokyo Woman's Christian University, Tokyo 167-8585, Japan
\bigskip
}
\maketitle
\begin{abstract}\noindent
When combining apparently inconsistent experimental results, one often implements errors on errors. The Particle Data Group's phenomenological prescription offers a practical solution but lacks a firm theoretical foundation. To address this, D'Agostini and Cowan have proposed Bayesian and frequentist approaches, respectively, both introducing gamma-distributed auxiliary variables to model uncertainty in quoted errors. In this Letter, we show that these two formulations admit a parameter-by-parameter correspondence, and are structurally equivalent. This identification clarifies how Bayesian prior choices can be interpreted in terms of frequentist sampling assumptions, providing a unified probabilistic framework for modeling uncertainty in quoted variances.
\end{abstract}

\newpage

\section{Introduction}

The combination of inconsistent experimental results is a recurring challenge in particle physics, particularly when managing and interpreting uncertainties. This issue arises when discrepancies between experimental data exceed the quoted statistical and systematic uncertainties, necessitating the implementation of additional uncertainty corrections.

The Particle Data Group (PDG) addresses this by introducing a phenomenological scale factor applied to the combined uncertainty:
\al{
S := \max\fn{1,\,\sqrt{\chi^2\ov N-1}},
}
where $N$ denotes the number of experimental results being combined and $\chi^2$ is the chi-squared value for the combination~\cite{PhysRevD.110.030001}. While this method is practical and widely used, it lacks a clear theoretical foundation, leaving room for more principled statistical treatments.

In Refs.~\cite{DAgostini:1999niu} and \cite{Cowan:2018lhq}, D’Agostini and Cowan independently proposed approaches to address the so-called ``errors on errors''---uncertainties in the quoted systematic uncertainties---based on Bayesian and frequentist inference, respectively. Both methods extend the standard profile likelihood by introducing auxiliary variables to model the uncertainty in the quoted errors, typically using gamma distributions.

While these ideas are individually known, the structural correspondence between the Bayesian and frequentist formulations has not been explicitly established in the literature.  
In this Letter, we show that the two approaches can in fact be mapped onto each other parameter by parameter. In particular, we demonstrate that a specific parameter choice in the Bayesian prior—often presented as an ad hoc recommendation—can be naturally understood as arising from a frequentist sampling model involving virtual repetitions of each experiment. This correspondence not only clarifies the origin of the so-called ``democratic skepticism'' advocated by D’Agostini, but also provides a concrete rationale for adopting symmetric gamma priors with shape and rate parameters set equal.

Our aim is not to introduce a new statistical model, but rather to clarify, connect, and unify existing approaches by revealing their structural equivalence.  
By identifying a parameter-by-parameter correspondence between the Bayesian and frequentist formulations, we provide a coherent probabilistic framework for modeling uncertainty in quoted variances.  
This perspective helps elucidate the rationale behind specific prior choices in Bayesian methods and may provide useful insight into empirical practices such as uncertainty rescaling in data combinations.

This Letter is organized as follows.  
In Sec.~\ref{D'Agostini section}, we review D'Agostini's Bayesian approach to incorporating errors on errors.  
In Sec.~\ref{Cowan section}, we present Cowan's frequentist formulation, recast in a form aligned with D'Agostini's approach.  
In Sec.~\ref{comparison section}, we compare the two and demonstrate their structural correspondence.  
In Sec.~\ref{summary and discussion section}, we summarize our findings and discuss their implications.




\section{D'Agostini's Bayesian approach}
\label{D'Agostini section}
We review D'Agostini's Bayesian approach for the errors on errors~\cite{DAgostini:1999niu}.

\subsection{Ordinary procedure derived from Bayesian inference}

Bayesian inference is based on the conditional probability\footnote{
Eq.~\eqref{Bayesian inference} follows from 
$P\fn{A\md B}={P\fn{B\md A}P\fn{A}\ov P\fn{B}}$, where $P\fn{A\md B}:={P\fn{A\cap B}\ov P\fn{B}}$.
}
\al{
f\fn{\tv\mu\md\ev{\bs\dD}}
	&\propto f\fn{\ev{\bs\dD}\md\tv\mu}f_\circ\fn{\tv\mu},
		\label{Bayesian inference}
}
where
\al{
\ev{\bs\dD}
	:=	\Set{\ev{\dD_i}}_{i=1,\dots,N}
	=	\Set{\ev{\dD_1}, \dots, \ev{\dD_N}}
}
is the data set, with roman indices $i, j, \dots$ running from $1$ to $N$ to label the experimental data, typically obtained from different experimental collaborations.  
Here, $f\fn{\ev{\bs\dD}\md\tv\mu}$ represents the likelihood of observing $\ev{\bs\dD}$ under the hypothesis that the true value is $\tv\mu$, and $f_\circ\fn{\tv\mu}$ is the \emph{prior} probability density function (PDF) of $\tv\mu$.  
In Eq.~\eqref{Bayesian inference}, both $f\fn{\tv\mu\md\ev{\bs\dD}}$ and $f_\circ\fn{\tv\mu}$ describe \emph{theoretical} distributions for the ``true'' value~$\tv\mu$, while $f\fn{\ev{\bs\dD}\md\tv\mu}$ represents the \emph{experimental} probability of obtaining the dataset $\ev{\bs\dD}$, assuming a true value $\tv\mu$.

\begin{table}[tp]
\centering
\begin{tabular}{ccc}
\hline
&Random variables & Fixed quantities\\
\hline
Theoretical&
$\trv{\h\mu}$, $\trv{\bh\sigma}$, $\trv{\bh r}$, $\trv{\bh\omega}$ &
$\tv\mu$, $\tv{\bs\sigma}$, $\tv{\bs r}$, $\tv{\bs\omega}$, $\tv{\bs v}$ \\
Experimental&
$\erv{\bh d}$, $\erv{\bh w}$, $\erv{\bh s}$ &
$\ev{\bs d}$, $\ev{\bs w}$, $\ev{\bs s}$\\
\hline
\end{tabular}
\caption{Variables used in this work, arranged by whether they are theoretical or experimental (rows) and whether they are random variables or fixed quantities (columns).  
Nuisance parameters are $\fv k$, $\fv\lambda$, $\fv n$, and $\fv\varepsilon$.}
\label{variables table}
\end{table}

In the usual frequentist approach, each experimental result $\erv{\h\dD_i}$ in the set
$\erv{\bs{\h\dD}} := \set{\erv{\h\dD_1},\dots,\erv{\h\dD_N}}$
is treated as a random variable independently distributed around the fixed true value $\tv\mu$.  
Here and hereafter, a hat $\hat{}\,$ denotes all estimators and other random variables derived from data, regardless of whether they appear in a frequentist or Bayesian framework.  
This notation distinguishes the hatted random variables from fixed parameters such as $\tv\mu$ and from deterministic values such as $\ev{\bs\dD}$.  
These distinctions are summarized in Table~\ref{variables table}.

Each $\erv{\h\dD_i}$ is typically modeled as following a normal distribution with mean $\tv\mu$ and standard deviation~$\tv{\sigma_i}$:
\al{
\mc N\fn{\ev{\dD_i}\md\tv\mu,\tv{\sigma_i^2}}
	:=	{1\ov\sqrt{2\pi}\tv{\sigma_i}}\exp\fnl{-{\pn{\ev{\dD_i}-\tv\mu}^2\ov2\tv{\sigma_i^2}}}.
		\label{normal distribution}
}
We also collectively write $\tv{\bs\sigma}:=\Set{\tv{\sigma_i}}_{i=1,\dots,N}$.  
Under the above assumption, the PDF for $\erv{\bs{\h\dD}}$ is given by
\al{
f\fn{\ev{\bs\dD}\md\tv\mu}
	=	f\fn{\ev{\bs\dD}\md\tv\mu,\tv{\bs\sigma}}
	:=	\prod_i\mc N\fn{\ev{\dD_i}\md\tv\mu,\tv{\sigma_i^2}}.
	\label{usual iid PDF}
}
Here and hereafter, we use the shorthand notation $\prod_i := \prod_{i=1}^N$, $\sum_i := \sum_{i=1}^N$, etc.

Now we come back to the Bayesian approach.
When one further assumes that the prior is uniform, $f_\circ\fn{\tv\mu}=\tx{const.}$,\footnote{
The uniform prior adopted here is improper and non-integrable, but this does not affect our subsequent analysis as long as the likelihood function~\eqref{posterior ordinary} decays sufficiently fast.
}
and that $\tv{\bs\sigma}$ are exactly equal to the \emph{quoted stated uncertainties} by the experimental collaborations,
\al{
\ev{\bs\sD}:=\Set{\ev{\sD_1},\dots,\ev{\sD_N}},
	\label{quoted stated uncertainty}
}
one ends up with the ordinary procedure:
\al{
f\fn{\tv\mu\md\ev{\bs\dD},\tv{\bs\sigma},f_\circ\fn{\tv\mu}=\tx{const.}}
	&\propto
		\prod_i\mc N\fn{\ev{\dD_i}\md\tv\mu,\ev{\sD_i^2}}
	\propto	\mc N\fn{\tv\mu\md\ev{\ol\dD},\ev{\ol\sD^2}},
	\label{posterior ordinary}
}
where
\al{
\ev{\ol\sD}
	&:=	{1\ov\sqrt{\sum_i{1\ov\ev{\sD_i^2}}}},&
\ev{\ol\dD}
	&:=	{\sum_i{\ev{\dD_i}\ov\ev{\sD_i^2}}\ov\sum_j{1\ov\ev{\sD_j^2}}}
	=	\ev{\ol\sD^2}\sum_i{\ev{\dD_i}\ov\ev{\sD_i^2}}.
}
Throughout this Letter, an overbar denotes an averaged quantity.
Accordingly, the expected value, the variance, and the standard deviation are
\al{
\E\fnl{\trv{\h\mu}}
	&:=	\int \df\tv\mu\,\tv\mu\,\mc N\fn{\tv\mu\md\ev{\ol\dD},\ev{\ol\sD^2}}
	=	\ev{\ol\dD},\\
\V\fnl{\trv{\h\mu}}
	&:=	\E\fnl{\pn{\trv{\h\mu}-\E\fnl{\trv{\h\mu}}}^2}
	=	\E\fnl{\trv{\h\mu^2}}-\E^2\fnl{\trv{\h\mu}}
	=	\ev{\ol\sD^2},\\
\sigma\fnl{\trv{\h\mu}}
	&:=	\sqrt{\V\fnl{\trv{\h\mu}}}
	=	\ev{\ol\sD},
}
for the normal distribution.

\subsection{Probabilistic modeling of skepticism}
In the above, we have first treated the standard deviations $\tv{\bs\sigma}$ as unknown quantities, and then assumed them to be fixed at $\ev{\bs\sD}$, namely, $\tv{\sigma_i}=\ev{\sD_i}$.
Therefore, it is natural to promote $\tv{\bs\sigma}$ to a set of random variables:
\al{
\trv{\bs{\h\sigma}}=\Set{\trv{\h\sigma_1},\dots,\trv{\h\sigma_N}}.
}

D'Agostini introduces the following ratios as alternative random variables to $\trv{\bs{\h\sigma}}$:
\al{
\trv{\h\rD_i}
	&:=	{\trv{\h\sigma_i}\ov\ev{\sD_i}}.
		\label{sigma vs sr}
}
The basic assumption is that the inverse squares of these ratios
\al{
\trv{\h\omega_i}
	&:=	{1\ov\trv{\h\rD_i^2}}
	=	{\ev{\sD_i^2}\ov\trv{\h\sigma_i^2}}
}
follow the \emph{gamma distribution}, whose PDF is given by\footnote{
We follow the standard convention in statistics, where the shape parameter is listed first, followed by the rate parameter. This contrasts with the convention used by the PDG~\cite{PhysRevD.110.030001}, where the rate parameter appears first.
}
\al{
g\fn{\tv{\omega_i}\md\fv{\kD},\fv{\lamD}}
	&:=	{\fv{\lamD^{\kD}}\ov\Gamma\fn{\fv{\kD}}}
		\tv{\omega_i^{\fv{\kD}-1}}e^{-\fv{\lamD}\tv{\omega_i}},
	\label{gamma distribution}
}
where $\fv{\kD}>0$ and $\fv{\lamD}>0$ are shape and rate parameters, respectively, which are treated as nuisance parameters.\footnote{\label{iid footnote}%
D'Agostini first introduces $i$-dependent $\fv{\kD_i}$ and $\fv{\lamD_i}$, and later assumes that $\trv{\h\omega_i}$ are identically distributed for all $i$, namely, $\fv{\kD_i} = \fv\kD$ and $\fv{\lamD_i} = \fv\lamD$, calling it \emph{democratic skepticism}.
We adopt this assumption from the beginning, since the $i$-dependence can be trivially recovered if needed.
}

The gamma distribution satisfies the following scaling property:
\al{
\alpha g\fn{\alpha\tv{\omega_i}\md\fv{\kD},\fv{\lamD}}
	&=	g\fn{\tv{\omega_i}\md\fv{\kD},\alpha\fv{\lamD}},
		\label{gamma scaling}
}
with $\alpha>0$.
For later convenience, we also list its expected value and variance:
\al{
\E\fnl{\trv{\h\omega_i}}
	&=	\int_0^\infty\df\tv{\omega_i}\,\tv{\omega_i}\,g\fn{\tv{\omega_i}\md\fv{\kD},\fv{\lamD}}
	=	{\fv{\kD}\ov\fv{\lamD}},
		\label{omega expected}\\
\V\fnl{\trv{\h\omega_i}}
	&=	\E\fnl{\pn{\trv{\h\omega_i}-\E\fnl{\trv{\h\omega_i}}}^2}
	=	\E\fnl{\trv{\h\omega_i^2}}-\E^2\fnl{\trv{\h\omega_i}}
	=	{\fv{\kD}\ov\fv{\lamD^2}}.
		\label{omega variance}
}
They both exist for all admissible $\fv\lamD>0$.

One of the main purposes of this Letter is to provide a clearer theoretical foundation for the somewhat arbitrary assumption~\eqref{gamma distribution} by examining how it aligns with Cowan's frequentist approach.

\subsection{D'Agostini's ratio}
Finally, taking into account the Jacobian in the probability element, we obtain the PDF in terms of the ratio:
\al{
f\fn{\tv{\rD_i}\md\fv{\kD},\fv{\lamD}}\df\tv{\rD_i}
	&=	g\fn{\tv{\omega_i}\md\fv{\kD},\fv{\lamD}}\df\tv{\omega_i}
	=	{2\ov\tv{\rD_i^3}}\,g\fn{{1\ov\tv{\rD_i^2}}\md\fv{\kD},\fv{\lamD}}\df\tv{\rD_i}
	=	{2\fv{\lamD^{\kD}} \ov\Gamma\fn{\fv{\kD}}}
		{e^{-\fv{\lamD}/\tv{\rD_i^2}}\ov\tv{\rD_i^{2\fv{\kD}+1}}}\df\tv{\rD_i}.
		\label{prior for f}
}
Consequently,
\al{
\E\fnl{\trv{\h\rD_i}}
	&=	\int_0^\infty\df\tv{\rD_i}\,\tv{\rD_i}\,f\fn{\tv{\rD_i}\md\fv{\kD},\fv{\lamD}}
	=	{\fv{\sqrt{\lamD}}\,\Gamma\fn{\fv{\kD}-{1\ov2}}\ov\Gamma\fn{\fv{\kD}}},
		\label{E D'Agostini}\\
\V\fnl{\trv{\h\rD_i}}
	&=	\fv{\lamD}\pn{{1\ov\fv{\kD}-1}-{\Gamma^2\fn{\fv{\kD}-{1\ov2}}\ov\Gamma^2\fn{\fv{\kD}}}}.
		\label{V D'Agostini}
}
They both exist when $\fv{\lamD}>0$ and $\fv{\kD}>1$.

\begin{figure}[tp]
\centering
\hfill
\includegraphics[width=0.4\textwidth]{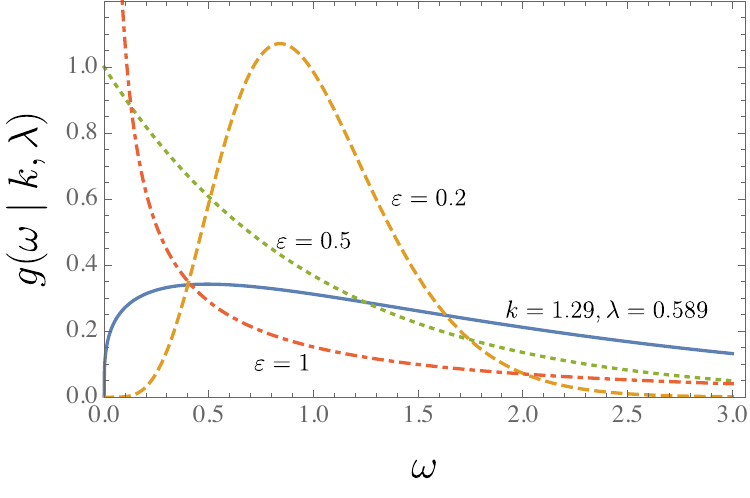}\hfill
\includegraphics[width=0.4\textwidth]{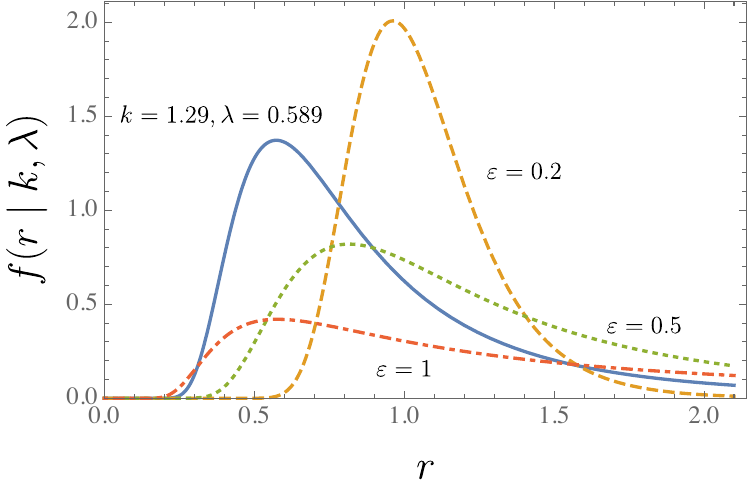}\hfill\mbox{}
\caption{
PDFs $g\fn{\tv\omega\md\fv\kD,\fv\lamD}$ and $f\fn{\tv r\md\fv\kD,\fv\lamD}$ are shown in the left and right panels, respectively. For simplicity, the $i$-dependence of $\tv{\omega_i}$ and $\tv{r_i}$ is omitted in this figure and caption.  
The nuisance parameters are set to D'Agostini's recommended values $\fv k = 1.29$, $\fv\lambda = 0.589$ (solid), and to Cowan's $\fv\kD = \fv\lamD = 1/4 \fv{\varepsilon^2}$ with $\fv\varepsilon = 0.2$ (dashed), $0.5$ (dotted), and $1$ (dot-dashed), corresponding to $\fv\kD = \fv\lamD = 6.25$, $1$, and $0.25$, respectively.  
The associated numbers of virtual repetitions are $\fv n = 13.5$, $3$, and $1.5$. Among these, the choice $\fv\varepsilon = 0.5$ corresponds to our version of D’Agostini’s criterion $\V\fnl{\trv{\h\omega}} = \E^2\fnl{\trv{\h\omega}}$.
}
\label{PDF figure}
\end{figure}

The above expressions~\eqref{E D'Agostini} and \eqref{V D'Agostini} can be used to choose natural values of $\fv{\lamD}$ and $\fv{\kD}$.  
Reasonable constraints might be
\al{
\E\fnl{\trv{\h\rD_i}}
	&=	1,&
\V\fnl{\trv{\h\rD_i}}
	&\simeq
		\E^2\fnl{\trv{\h\rD_i}}
	=	1.
		\label{expectation of r in D'Agostini}
}
The former condition follows from the assumption that the standard deviations on average agree with the stated uncertainties, while the latter provides a conservative estimate.\footnote{
As discussed in Section 2.3.6 of Ref.~\cite{Aoyama:2020ynm}, some published measurements may contain biases or underestimate uncertainties, potentially affecting D'Agostini's assumption that quoted uncertainties are unbiased and conservatively estimated.
}
The constraints~\eqref{expectation of r in D'Agostini} are satisfied by  
\al{
\fv{\kD}&\simeq 1.29,&
\fv{\lamD}&\simeq 0.589.
\label{recommended values}
}
These are the recommended values by D'Agostini; see Fig.~\ref{PDF figure} for the corresponding PDFs.

We see that the former condition in Eq.~\eqref{expectation of r in D'Agostini} is physically motivated, reflecting the assumption that the quoted errors are unbiased, whereas the latter serves as a conservative estimate with no clear theoretical preference.  
As an alternative to $\E\fnl{\trv{\h\rD_i}} = 1$, one may consider imposing the condition
\al{
\E\fnl{\trv{\h\omega_i}} = 1.
}
Together with Eq.~\eqref{omega expected}, this leads to
\al{
\fv\kD = \fv\lamD,
	\label{our recommendation}
}
with the variance
\al{
\V\fnl{\trv{\h\omega_i}} = \frac{1}{\fv\kD}.
}
We will show in Section~\ref{comparison section} that this choice arises naturally in Cowan's frequentist formulation, offering a compelling theoretical perspective.  

Further imposing the conservative estimate $\V\fnl{\trv{\h\omega_i}} = 1$ yields
\al{
\fv\kD = \fv\lamD = 1,
	\label{further recommendation}
}
for which D'Agostini's ratio satisfies $\E\fnl{\trv{\h r_i}} = \sqrt\pi \simeq 1.77$, while its variance diverges as $\V\fnl{\trv{\h r_i}} \to 1 / (\fv\kD - 1) \to \infty$.

\subsection{Likelihood function}
For the likelihood function $f\fn{\ev{\bs\dD}\md\tv\mu}$ in the Bayesian inference~\eqref{Bayesian inference},
we assume that the results are independent from each other and further
that $\erv{\h\dD_i}$ follows a normal distribution with standard deviation $\tv{\rD_i}\ev{\sD_i}$ ($=\tv{\sigma_i}$ in the previous notation):
\al{
f\fn{\ev{\bs\dD}\md\tv\mu}
	=	\prod_i f\fn{\ev{\dD_i}\md\tv\mu,\ev{\sD_i},\fv{\kD},\fv{\lamD}},
}
where
\al{
f\fn{\ev{\dD_i}\md\tv\mu,\ev{\sD_i},\fv{\kD},\fv{\lamD}}
	&=	\int_0^\infty \df\tv{\rD_i}\,
		\mc N\fn{\ev{\dD_i}\md\tv\mu,\pn{\tv{\rD_i}\ev{\sD_i}}^2}
		f\fn{\tv{\rD_i}\md\fv{\kD},\fv{\lamD}};
		\label{integrated likelihood for f}
}
cf.\ Eq.~\eqref{usual iid PDF}.
Putting Eqs.~\eqref{normal distribution} and \eqref{prior for f}, this assumption leads to
\al{
f\fn{\ev{\dD_i}\md\tv\mu,\ev{\sD_i},\fv{\kD},\fv{\lamD}}
	&=	{\fv{\lamD^{\kD}}\ov\sqrt{2\pi}\ev{\sD_i}}{\Gamma\fn{\fv{\kD}+{1\ov2}}\ov\Gamma\fn{\fv{\kD}}}
		\pn{\fv{\lamD}+{\pn{\ev{\dD_i}-\tv\mu}^2\ov2\ev{\sD_i^2}}}^{-\pn{\fv{\kD}+{1\ov2}}}.
		\label{t-distribution}
}
This PDF exhibits the kernel of a Student’s $t$~distribution, reflecting marginalization over a gamma-distributed variance.\footnote{
Integrating the normal likelihood in Eq.~\eqref{integrated likelihood for f} over the gamma prior yields a Student’s $t$ distribution with $\nu = 2\fv\kD$. Equation~\eqref{t-distribution} matches its kernel up to reparameterization, exposing the heavy tails from variance uncertainty.
}

When one further assumes a uniform prior distribution for $\trv{\h\mu}$, the resulting posterior takes the form
\al{
f\fn{\tv\mu\md\ev{\bs\dD},\ev{\bs\sD},\fv\kD,\fv\lamD}
	\propto
		f\fn{\ev{\bs\dD}\md\tv\mu,\ev{\bs\sD},\fv\kD,\fv\lamD}
	\propto
		\prod_i\pn{\fv\lamD+{\pn{\ev{\dD_i}-\tv\mu}^2\ov2\ev{\sD_i^2}}}^{-\pn{\fv\kD+{1\ov2}}}.
		\label{final practical form of f}
}
Finally, integrating out the nuisance parameters $\fv\kD$ and $\fv\lamD$, we obtain
\al{
f\fn{\tv\mu\md\ev{\bs\dD},\ev{\bs\sD}}
	&=	\int\df\fv\kD\,\df\fv\lamD\,f\fn{\fv\kD,\fv\lamD}f\fn{\tv\mu\md\ev{\bs\dD},\ev{\bs\sD},\fv\kD,\fv\lamD},
}
where $f\fn{\fv\kD,\fv\lamD}$ quantifies the confidence in each possible set of parameters.
Alternatively, Eq.~\eqref{final practical form of f} can be evaluated simply by substituting D'Agostini's recommended values~\eqref{recommended values} or our own~\eqref{our recommendation}, which is arguably the only feasible approach in practice.


%

\section{Cowan's frequentist approach}\label{Cowan section}

We review Cowan's frequentist approach to dealing with errors on errors~\cite{Cowan:2018lhq}.

\subsection{Cowan’s sampling scheme}
In the Bayesian approach, the ``true value'' $\trv{\h\mu}$ is treated as a \emph{theoretical} random variable that follows a prior distribution $f_\circ\fn{\tv\mu}$ and is updated to a posterior distribution $f\fn{\tv\mu\md\ev{\bs\dD},\ev{\bs\sD}}$ upon observing data.
In contrast, the frequentist approach considers $\tv\mu$ to be the fixed ``true value'', while the observed variables fluctuate around it according to Eq.~\eqref{usual iid PDF}.

To handle the ``errors on errors'' within the frequentist framework, Cowan provides a way to treat each estimated variance $\erv{\h w_i}$ as an \emph{experimental} random variable fluctuating around a so-to-say ``true variance'' $\tv{v_i}$.
To implement this idea, Cowan introduces $\fv n$ (virtual) repetitions for each real experiment $i$, resulting in $\fv n$ random variables for each:\footnote{\label{iid footnote frequentist}%
In principle, $\fv n$ may vary across experiments, resulting in an $i$-dependent $\fv{n_i}$. However, this dependence is typically neglected in practice~\cite{Cowan:2021sdy,Canonero:2024kzk}, corresponding to D'Agostini's notion of democratic skepticism; see footnote~\ref{iid footnote}. Here we also omit the $i$-dependence, as it can be trivially recovered if needed.
}
\al{
\Set{\erv{\h d_{i\,\tf j}}}_{\tf j=1,\dots,\fv n}.
}
These are assumed to follow normal distributions:
\al{
\mc N\fn{\ev{\dD_i}\md\tv\mu,\tv{v_i}}.
	\label{normal distribution for each}
}
Here and hereafter, the Roman typewriter-style index $\tf j$ runs from 1 to $\fv n$, indexing the virtual repetitions associated with each real experiment $i$.
The sample mean and the unbiased sample variance of these outcomes are defined as usual:
\al{
\erv{\ol{\h d_i}}
	&:=	{1\ov\fv n}\sum_{\tf j=1}^{\fv n}\erv{\h d_{i\,\tf j}},&
\erv{\h w_i}
	&:=	{1\ov \fv n-1}\sum_{\tf j=1}^{\fv n}\pn{\erv{\h d_{i\,\tf j}}-\erv{\ol{\h d_i}}}^2.
}

Given the normal distributions~\eqref{normal distribution for each}, the random variable
$
{\fv n-1\ov\tv{v_i}}\erv{\h w_i}
$
follows a chi-squared distribution with $\fv n-1$ degrees of freedom, which then can be written in terms of the gamma distribution~\eqref{gamma distribution} as:
\al{
\chi^2_{\fv n-1}\fn{{\fv n-1\ov\tv{v_i}}\ev{w_i}}
	&=	g\fn{{\fv n-1\ov\tv{v_i}}\ev{w_i}\md{\fv n-1\ov2},{1\ov2}}.
		\label{chi squared vs gamma}
}
That is, the PDF for the variance is obtained from the probability elements as
\al{
p\fn{\ev{w_i}\md\tv{v_i},\fv n}\df\ev{w_i}
	&=	g\fn{{\fv n-1\ov\tv{v_i}}\ev{w_i}\md{\fv n-1\ov2},{1\ov2}}\df\fn{{\fv n-1\ov\tv{v_i}}\ev{w_i}}\nn
	&=	g\fn{\ev{w_i}\md{\fv n-1\ov2},{\fv n-1\ov2\tv{v_i}}}\df\ev{w_i},
		\label{gamma P}
}
where we used the scaling property~\eqref{gamma scaling}.
It follows that
\al{
\E\fnl{\erv{\h w_i}}
	&=	{\tv{v_i}},&
\V\fnl{\erv{\h w_i}}
	&=	{2\ov\fv n-1}\tv{v_i^2},&
\sigma\fnl{\erv{\h w_i}}
	&=	\sqrt{2\ov\fv n-1}\,\tv{v_i}.
}

\subsection{Errors-on-errors parameter}
Cowan defines what is called the errors-on-errors parameter:\footnote{
We follow the notation $\fv\varepsilon$ used in Ref.~\cite{Canonero:2023sua}, rather than $\fv r$ in Ref.~\cite{Cowan:2018lhq}, for the same quantity (not to be confused with D'Agostini's $\tv{\rD_i}$). Although this parameter may in principle depend on $i$, we omit the $i$-dependence while retaining the original name, ``errors-on-errors,'' which reflects its possible variability across experiments; see also footnotes~\ref{iid footnote} and \ref{iid footnote frequentist}.
}
\al{
\fv\varepsilon
	&:=	{1\ov2}{\sigma\fnl{\erv{\h w_i}}\ov\E\fnl{\erv{\h w_i}}}
	=	{1\ov\sqrt{2\pn{\fv n-1}}}.
	\label{errors on errors}
}
The spirit behind this definition can be explained as follows~\cite{Cowan:2018lhq}:
For $\erv{\h s_i^2}:=\erv{\h w_i}$, the standard error propagation gives
$\Delta\ev{w_i} \simeq 2\ev{s_i}\Delta\ev{s_i}$.
Cowan then expects the analogous relation between the standard deviations:
$\sigma\fnl{\erv{\h w_i}} \simeq 2\E\fnl{\erv{\h s_i}}\sigma\fnl{\erv{\h s_i}}$.
If one were to approximate
$\E\fnl{\erv{\h w_i}} \sim \E^2\fnl{\erv{\h s_i}}$,
then one would vaguely expect\footnote{
Note, however, that this naive treatment leads to $\V\fnl{\erv{\h s_i}} \sim 0$, implying $\sigma\fnl{\erv{\h s_i}} \sim 0$.  
This limitation can also be seen more explicitly: It suggests that
$
{1\ov2}{\sigma\fnl{\erv{\h w_i}}\ov\E\fnl{\erv{\h w_i}}}
	=	\frac{1}{2}\sqrt{\frac{\Braket{\erv{\h s_i^4}}}{\Braket{\erv{\h s_i^2}}^2}-1}
$
should be close to
$
{\sigma\fnl{\erv{\h s_i}}\ov\E\fnl{\erv{\h s_i}}}
	=	\sqrt{\frac{\Braket{\erv{\h s_i^2}}}{\Braket{\erv{\h s_i}}^2}-1},
$
which is unjustifiable.
}
\al{
\fv\varepsilon
	\sim	{\sigma\fnl{\erv{\h s_i}}\ov\E\fnl{\erv{\h s_i}}}.
	\label{meaning of Cowan's r}
}
In this sense, $\fv\varepsilon$ is said to be a measure of relative error on error.

Regardless of how it was initially motivated, the errors-on-errors parameter~\eqref{errors on errors} serves as a convenient tool to express the auxiliary quantity $\fv n$ in terms of an intuitive measure of relative uncertainty:
\al{
\fv n
	&=	1+\frac{1}{2\fv{\varepsilon^2}}.
}
The resultant PDF reads
\al{
p\fn{\ev{w_i}\md\tv{v_i},1+{1\ov2\fv{\varepsilon^2}}}
	&=	g\fn{\ev{w_i}\md{1\ov4\fv{\varepsilon^2}},{1\ov4\fv{\varepsilon^2}\tv{v_i}}},
	\label{P v sigma}
}
and the variance and standard deviation of $\erv{\h w_i}$ become
\al{
\V\fnl{\erv{\h w_i}}
	&=	4\fv{\varepsilon^2}\tv{v_i^2},&
\sigma\fnl{\erv{\h w_i}}
	=	\sqrt{\V\fnl{\erv{\h w_i}}}
	&=	2\fv\varepsilon\tv{v_i}.
}

\section{Comparison between D'Agostini and Cowan}\label{comparison section}
This section presents the main findings of the Letter. We compare D'Agostini's Bayesian and Cowan's frequentist approaches to uncertainty in quoted variances, focusing on the structure of their respective probabilistic formulations.

D'Agostini's Bayesian approach treats $\ev{s_i^2}$ and $\trv{\h\sigma_i^2}$ as the experimental deterministic value and the \emph{theoretical} random variable, respectively, whereas Cowan's frequentist approach treats $\erv{\h w_i}$ and $\tv{v_i}$ as the \emph{experimental} random variable and the deterministic theoretical value.

Here we point out, on physical grounds, that the ratio $\ev{s_i^2}/\trv{\h\sigma_i^2}$ corresponds to the ratio $\erv{\h w_i}/\tv{v_i}$:
\al{
\underset{\tx{D'Agostini's Bayesian}}{\trv{\h\omega_i}={1\ov\trv{\h r_i^2}}={\ev{s_i^2}\ov\trv{\h\sigma_i^2}}}
	&\sr{\tx{correspond}}\longleftrightarrow
		\underset{\tx{Cowan's frequentist}}{\erv{\h w_i}\ov\tv{v_i}}.
}
In Cowan's Eq.~\eqref{chi squared vs gamma} or \eqref{gamma P}, the probability element reads
\al{
\chi^2_{\fv n-1}\fn{{\fv n-1\ov\tv{v_i}}\ev{w_i}}
\df\fn{{\fv n-1\ov\tv{v_i}}\ev{w_i}}
	&=	g\fn{{\ev{w_i}\ov\tv{v_i}}\md{\fv n-1\ov2},{\fv n-1\ov2}}\df{\ev{w_i}\ov\tv{v_i}},
}
where we have again used the scaling property~\eqref{gamma scaling}.  
This can be directly compared with D'Agostini's gamma distribution in Eq.~\eqref{gamma distribution}:
\al{
g\fn{\tv{\omega_i}\md\fv{\kD},\fv{\lamD}}\df\tv{\omega_i}.
}
The shape and rate parameters can thus be identified as
\al{
\fv\kD = \fv\lamD = {\fv n - 1 \ov 2} = {1 \ov 4 \fv{\varepsilon^2}}.
}

We see that the choice $\fv\kD=\fv\lamD$, as suggested in Eq.~\eqref{our recommendation}, provides a better match with the result from Cowan's frequentist approach.\footnote{
Cowan also touches on related Bayesian approaches in his talk~\cite{CowanTalkCERN}, noting that choosing $\fv\kD = \fv\lamD$ yields a gamma distribution with unit expectation and adjustable variance.  
His reference to the inverse gamma distribution, in comparison with D'Agostini's approach, suggests a perspective not fully aligned with the formulation considered here, which relies exclusively on the gamma distribution.
}
The corresponding PDFs are shown in Fig.~\ref{PDF figure} with the values $\fv\varepsilon=0.2$, $0.5$, and $1$, namely, $\fv\kD=\fv\lamD=6.25$, $1$, and $0.25$, respectively.  
Accordingly, the corresponding numbers of virtual repetitions are $\fv n = 13.5$, $3$, and $1.5$, respectively.  
Among these, the choice $\fv\varepsilon = 0.5$ corresponds to our version of D'Agostini's criterion $\V\fnl{\trv{\h\omega_i}} = \E^2\fnl{\trv{\h\omega_i}}$, leading to Eq.~\eqref{further recommendation}.

\section{Summary and discussion}\label{summary and discussion section}

In this Letter, we have clarified the structural correspondence between D'Agostini's Bayesian and Cowan's frequentist approaches to implementing errors on errors. Both methods enhance the conventional treatment of experimental uncertainties by modeling auxiliary random variables with gamma distributions. Although their starting points differ---Bayesian priors versus frequentist sampling models---we have established a parameter-by-parameter mapping between them.

In particular, we have demonstrated that the Bayesian assumption of a symmetric gamma prior with shape and rate parameters $\pn{\fv\kD, \fv\lamD}$ corresponds naturally to Cowan's frequentist model with $\fv n$ virtual repetitions, where $\fv\kD = \fv\lamD = \pn{\fv n-1}/2 = 1/4\fv{\varepsilon^2}$. This identification provides a theoretical rationale for the prior choice $\fv\kD = \fv\lamD$, ensuring $\E\fnl{\trv{\h\omega_i}} = 1$ and $\V\fnl{\trv{\h\omega_i}}= 1/\fv\kD$, and provides a basis for choosing specific values such as $\fv\kD = \fv\lamD = 1$, which satisfy $\V\fnl{\trv{\h\omega_i}} = \E^2\fnl{\trv{\h\omega_i}}$ as our version of D’Agostini's recommendation.

Our aim has not been to propose a new statistical model, but rather to unify existing approaches by revealing their underlying equivalence. This correspondence offers a principled perspective that sheds light on empirical procedures such as uncertainty rescaling, and clarifies how specific prior choices in Bayesian methods can be interpreted in terms of frequentist assumptions about sampling variability.

We hope that the parameter-by-parameter identification presented here serves to bridge the conceptual gap between Bayesian and frequentist perspectives, and provides a coherent probabilistic framework that may serve as a guiding principle for future work involving uncertainty modeling and data combination strategies.

Before closing, we note possible directions for future work. The hierarchical Bayesian model with hyperpriors on variance parameters~\cite{Erler:2020bif} has been proposed as a way to address issues like ``sub-sampling''~\cite{DAgostini:2020pim}, where iterative rescaling can bias the combined central value. It would be worthwhile to examine whether this issue can also be resolved within a frequentist approach, and whether the correspondence identified in this Letter continues to hold in that context. In parallel, it would also be valuable to explore how the frameworks discussed here perform when applied to real-world data sets~\cite{Crivellin:2023zui}, such as those underlying the $W$ boson mass~\cite{Rallapalli:2023ptp}, the muon $g-2$ anomaly~\cite{Aoyama:2020ynm},\footnote{
Applications to the muon $g-2$ have already been considered in Ref.~\cite{Cowan:2021sdy}; see also recent developments on the data-theory comparison in Refs.~\cite{Aliberti:2025beg,Muong-2:2025xyk}.
}
the neutron lifetime~\cite{Fuwa:2024cdf} (and references therein), and the $\sigma_8$ tension in cosmology~\cite{Karim:2024luk}. Such applications may offer not only practical validation but also a clearer understanding of the statistical structure behind persistent discrepancies in precision measurements.

\clearpage
\bibliographystyle{JHEP}
\bibliography{refs}
\end{document}